\documentclass{article}
\usepackage{arxiv}
\usepackage[utf8]{inputenc} 
\usepackage[T1]{fontenc}    
\usepackage{hyperref}       
\usepackage{url}            
\usepackage{booktabs}       
\usepackage{amsfonts}       
\usepackage{nicefrac}       
\usepackage{microtype}      
\usepackage{lipsum}		
\usepackage{graphicx}
\usepackage{doi}
\usepackage{tabularx}

\usepackage{color}
\usepackage{tikz}
\usetikzlibrary{quantikz}
\usepackage{amssymb} 
\usepackage{caption}
\usepackage{subcaption}
\usepackage{tabularx}
\usepackage{ragged2e}

\newcommand{\Ry}{\text{R}$_\text{y}$}
\newcommand{\kolkata}{\texttt{ibmq\_kolkata}}
\newcommand{\ehningen}{\texttt{ibmq\_ehningen}}
\newcommand{\lima}{\texttt{ibmq\_lima}}
\newcommand{\qasm}{\texttt{qasm\_simulator}}
\newcommand{\ncalls}{N_{\text{calls}}}
\newcommand{\nforward}{N_{\text{forward}}}
\newcommand{\nbackward}{N_{\text{backward}}}
\newcommand{\expect}[1]{ \langle #1 \rangle} 

\title{Hybrid quantum transfer learning for crack image classification on NISQ hardware}

\author{ \href{https://orcid.org/0000-0002-9955-4809}{\includegraphics[scale=0.06]{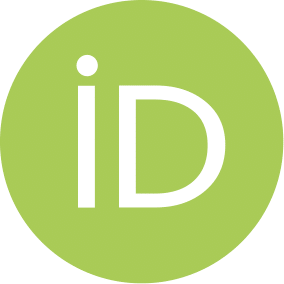}\hspace{1mm}Alexander Geng}\thanks{Corresponding author} \\
	Fraunhofer Institute for Industrial Mathematics ITWM\\ 
	Fraunhofer-Platz 1, 67663 Kaiserslautern\\            
	\texttt{alexander.geng@itwm.fraunhofer.de}\\
	\And
	\href{https://orcid.org/0000-0001-9126-3495}{\includegraphics[scale=0.06]{orcid.png}\hspace{1mm}Ali Moghiseh}\\
	Fraunhofer Institute for Industrial Mathematics ITWM\\ 
	Fraunhofer-Platz 1, 67663 Kaiserslautern\\            
	\texttt{ali.moghiseh@itwm.fraunhofer.de}\\
	\And
	\href{https://orcid.org/0000-0002-8030-069X}{\includegraphics[scale=0.06]{orcid.png}\hspace{1mm}Claudia Redenbach}\\
	University of Kaiserslautern\\ 
	Gottlieb-Daimler-Straße 47, 67663 Kaiserslautern\\            
	\texttt{redenbach@rptu.de}\\
	\And
	\href{https://orcid.org/0000-0003-4903-3180}{\includegraphics[scale=0.06]{orcid.png}\hspace{1mm}Katja Schladitz} \\
	Fraunhofer Institute for Industrial Mathematics ITWM\\ 
	Fraunhofer-Platz 1, 67663 Kaiserslautern\\            
	\texttt{katja.schladitz@itwm.fraunhofer.de}\\}

\begin{document}

\maketitle

\begin{abstract}
    Quantum computers possess the potential to process data using a remarkably reduced number of qubits compared to conventional bits, as per theoretical foundations. However, recent experiments \cite{geng_improved_2023} have indicated that the practical feasibility of retrieving an image from its quantum encoded version is currently limited to very small image sizes. Despite this constraint, variational quantum machine learning algorithms can still be employed in the current noisy intermediate scale quantum (NISQ) era. An example is a hybrid quantum machine learning approach for edge detection \cite{geng_hybrid_2022}. In our study, we present an application of quantum transfer learning for detecting cracks in gray value images. We compare the performance and training time of PennyLane's standard qubits with IBM's qasm\_simulator and real backends, offering insights into their execution efficiency.
\end{abstract}


\section{\label{sec:Introduction}Introduction}
Image based crack detection in concrete is a fundamental technique for monitoring buildings like houses or bridges. A particular focus is on early crack detection when crack thickness is not more than one pixel such that the crack is hardly visible in the image.
We solved this task by using classical deep learning methods \cite{muller_application_2018}.

Quantum computing promises many advantages. Superposition refers to the ability of a quantum system to exist in multiple states simultaneously. Moreover, two or more qubits can be entangled and affect each other regardless of the distance between them. This allows us to change the state of a qubit without applying operations to it. Furthermore, quantum computers can perform multiple computations simultaneously across different quantum states, leveraging the principles of superposition and entanglement. Another useful property of quantum computing is the creation of a larger search space, which gives us, for example, the possibility to separate data points or features in a higher dimensional space \cite{nielsen_quantum_2010,bergholm_pennylane_2022}. 

However, the current noisy intermediate scale quantum (NISQ) era still poses a lot of challenges like hardware errors and the fact that quantum states cannot be measured without collapsing them (Copenhagen interpretation \cite{faye_copenhagen_2019}). A consequence of the latter is that an estimate of a qubit's quantum state after running a quantum circuit can only be derived from the frequencies of measuring the basis states 0 and 1 after repeated runs of the circuit.
This is a particular bottleneck when combining machine learning with quantum computing algorithms (quantum machine learning). In the training phase, multiple runs are needed to calculate the gradients and to update the parameters as in the classical case. We cannot store intermediate results that are needed, for example, for backpropagation \cite{izaac_quantum_2020}.

This paper shows how to deal with these problems in a specific use case. We review the current possibilities of how to update parameters in a quantum machine learning setting. Taking the limitations of current quantum hardware into account, we propose a quantum transfer learning method for crack classification using PennyLane software \cite{bergholm_pennylane_2022} and IBM's quantum computers \cite{ibm}.

The paper is organized as follows. Section~\ref{sec:basics} provides some basics of quantum computing and provides information on the software and hardware used. In Section~\ref{sec:Differentiation methods}, we describe the currently available differentiation methods. We explain our method for classifying cracks in Section~\ref{sec:method}. In Section~\ref{sec:Application and results}, experiments for a given image dataset and the results for three differentiation methods on PennyLane's simulator are reported. Additionally, we run the approach on IBM's current quantum computers. Section~\ref{sec:flexibility} describes modifications of our algorithm with results on PennyLane's simulator and Section~\ref{sec:Conclusion and Outlook} concludes the paper and gives a short outlook.

\section{\label{sec:basics}Quantum computing basics, software, and hardware}

Before we discuss quantum hardware and crack classification methods, we summarize some basic concepts of quantum computing \cite{nielsen_quantum_2010}. Quantum computing follows different laws than classical computing. The difference starts with the basic elements. Classically, we have so-called bits, which can be either 0 or 1. The quantum analogues are quantum bits (qubits) -- two-state quantum systems that allow for more flexibility. Analogously to 0 and 1, there are two basis states of a qubit: $\ket{0}=(1,0)^T$ and $\ket{1}=(0,1)^T$. However, any linear combination (superposition)
\begin{equation}\label{geng_alexander:equation_qubit}
    \ket{\psi}=\alpha \ket{0}+\beta \ket{1},
\end{equation}
of the basis states with $\alpha, \beta \in \mathbb{C}$ and $|\alpha|^2+|\beta|^2=1$ defines a possible state, too. The overall phase of a quantum state is unobservable \cite{nielsen_quantum_2010}. That is, $\ket{\psi}$ and $e^{i\xi} \ket{\psi}$ for $\xi \in [0, 2 \pi]$ define the same state. 
Therefore, it is sufficient to consider $\alpha \in \mathbb{R}$. 

\begin{figure}[tb]
    \centering
    \includegraphics[width=0.3\textwidth]{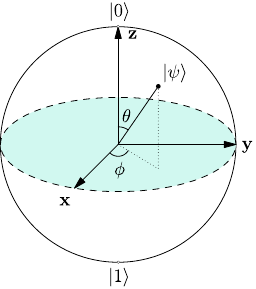}
    \caption{Visualization of a quantum state $\ket{\psi}$ in the Bloch sphere.}
    \label{fig:geng_alexander:bloch}
\end{figure}

As a consequence, the state of a single qubit can be visualized as a point on the unit sphere in $\mathbb{R}^3$ (Bloch sphere) with spherical coordinates $\phi$ and $\theta$, where $\alpha=\cos(\theta/2)$ and $\beta=e^{i\phi}\sin(\theta/2)$. Figure~\ref{fig:geng_alexander:bloch} shows the Bloch sphere with the spherical coordinates.

All operations on a qubit must preserve the condition $|\alpha|^2+|\beta|^2=1$, and thus can be represented by $2\times 2$ unitary matrices.
Standard operations (so-called gates) acting on single qubits are
\begin{equation}\label{equ:gates}
    X=\left(\begin{array}{ccc}
        0 && 1\\
        1 && 0
    \end{array}\right), \quad
    H=\frac{1}{\sqrt{2}}\left(\begin{array}{ccc}
        1&& 1\\
        1 && -1
    \end{array}\right), \quad
    R_y(\theta)=\left(\begin{array}{ccr}
        \cos(\theta/2) && -\sin(\theta/2) \\
        \sin(\theta/2) && \cos(\theta/2)
    \end{array}\right),
\end{equation}
where the X-gate acts like a classical NOT operator and the Hadamard gate (H) superposes the basic states of a single qubit. The single-qubit rotation gate (\Ry) rotates by $\theta$ about the y-axis of the Bloch sphere. In imaging applications, rotation gates can be used to encode gray values.

Additionally, we need operations that link two or more qubits. The most common operation in quantum computing is the controlled NOT-gate (CX-gate) taking two input qubits. The target qubit's state is changed depending on the state of the control qubit:
\begin{equation}\label{equ:cx_gates}
    \centering
    CX=\left(\begin{array}{ccccc}
        1 & 0 &0 & 0\\
        0 & 1 &0 & 0\\
        0 & 0 &0 & 1\\
        0 & 0 &1 & 0
        \end{array}\right).
\end{equation}
That means, if the control qubit is in state $\ket{1}$, we apply an X-gate to the target qubit. Otherwise, we do nothing. For example, assume our two-qubit system has the state $\ket{10}=\ket{1}\otimes\ket{0}$, where the first qubit is the control, the second the target qubit, and $\otimes$ is the tensor product. Then, the application of the CX-gate results in the state 
\begin{equation}
    \ket{11}=\ket{1}\otimes\ket{1}=(0, 0, 0, 1)^T.
\end{equation}
So basically, the application of quantum gates can be formulated in terms of linear algebra.

In general, we can apply any unitary operation to the target qubit. For example, a controlled rotation gate around the y-axis applies an \Ry-gate to the target qubit if and only if the control qubit is in state $\ket{1}$. We can also increase the number of control qubits even further. 

Applying such controlled operations to two or more qubits with the control qubits in superposition results in entanglement of the qubits involved. In terms of linear algebra, an entangled state of several qubits cannot be written as a tensor product of the states of the individual qubits. Entanglement is exactly where we benefit from the quantum computing properties. Together with superposition, entanglement allows using a logarithmically lower number of qubits compared to the number of classical bits.

\begin{figure}[tb]
    \begin{subfigure}[tb]{.7\linewidth}
        \includegraphics[width=\textwidth]{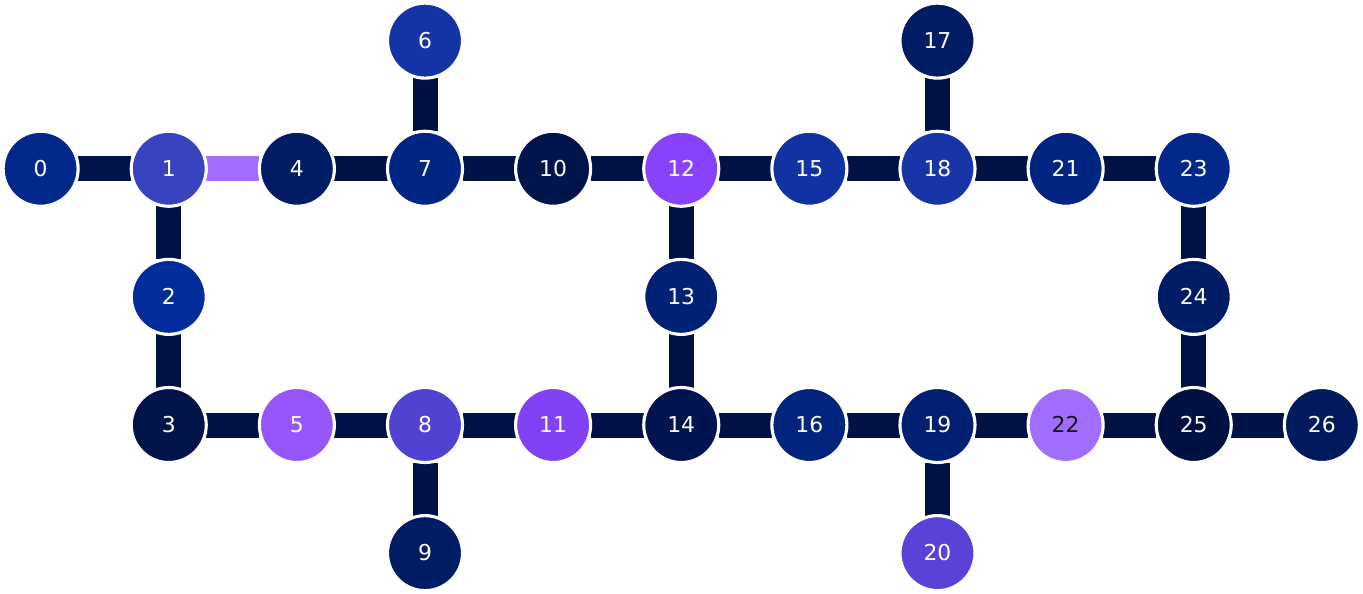}
        \caption{\ehningen.}
        \label{fig:geng_alexander:ehningen}
    \end{subfigure}
    \begin{subfigure}[tb]{.18\linewidth}
        \includegraphics[width=\textwidth]{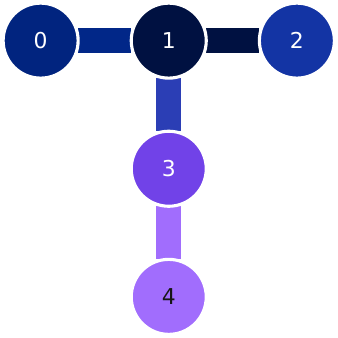}
        \caption{\lima.}
        \label{fig:geng_alexander:lima}
    \end{subfigure}
	\caption{Coupling maps of the backends used in this paper. Every circle represents a qubit, and lines represent connections between the qubits. Colors code the readout errors (circles) and the CX-errors for the connections (lines). Dark blue indicates a small error, and purple a large one. Errors are shown for \ehningen\, and \lima. IBM's backend \kolkata\, has the same coupling map as \ehningen, but actual performance errors differ slightly (see Tables~\ref{tab:geng_alexander:actual_performance} and \ref{tab:geng_alexander:error_rates}). }
	\label{fig:geng_alexander:used_backends}
\end{figure}

While all bits are connected to each other in classical computers, the qubits in IBM's quantum computers \cite{ibm}
are arranged in a special, the so-called heavy-hexagonal scheme (see the honeycomb structure in Figure~\ref{fig:geng_alexander:used_backends}). That is, each qubit is directly connected
to at most three other qubits. To apply two-qubit gates to unconnected qubits, the information has to be swapped to neighboring qubits by the application of additional CX-gates. Each CX-gate, however, increases the overall error considerably such that an algorithm should employ as few CX-gates as possible.


\begin{table}[tb]
    \caption{Processor type and actual performance of the used backends as measured in June 2023. The term quantum volume (QV) represents the quality of the backend. CLOPS is the number of circuit layer operations per second and gives a measure for the speed of the backend.}
    \label{tab:geng_alexander:actual_performance}
    \centering
    \begin{tabular}{lclccc}
        \hline\noalign{\smallskip}
        Backend &&Processor type& \# qubits & QV & CLOPS\\
        \noalign{\smallskip}\hline\noalign{\smallskip}
        \kolkata      &&Falcon r5& \phantom{1}27    & \phantom{1}128 & 2,000\\
        \ehningen    &&Falcon r5& \phantom{1}27    & \phantom{1}64 & 1,900\\
        \lima    &&Falcon r4T& \phantom{2}5    & \phantom{2}8 & 2,700\\
        \noalign{\smallskip}\hline
    \end{tabular}
\end{table}

\begin{table}[tb]
    \caption{Typical average calibration data of the three chosen backends. The values are from June 2023.}
    \label{tab:geng_alexander:error_rates}
    \centering
    \begin{tabular}{lcccccc}
        \hline\noalign{\smallskip}
        Backend & CX-error & Single qubit gate error & Readout error & T1 && T2 \\
        &in \% & in \% & in \% & 
        in \textmu s && in \textmu s\\
        \noalign{\smallskip}\hline\noalign{\smallskip}
        \kolkata         & 0.835 & 0.021 & 1.000 & 122.23    && \phantom{1}64.40 \\
        \ehningen         & 0.740 & 0.024 & 0.880 & 172.28    && 172.45 \\
        \lima         & 1.097 & 0.048 & 2.580 & 75.00    && 103.41 \\
        \noalign{\smallskip}\hline
    \end{tabular}
\end{table}

Lastly, the readout is also completely different for classical and quantum computing. On classical computers, one can always read the current state of the bits, copy them, or just continue running an algorithm with the same state of the bits as before the readout. Unfortunately, this is not possible on quantum computers. First, according to the no-cloning theorem \cite{nielsen_quantum_2010}, a state cannot be copied. Second, when measuring (reading out the state of) a qubit, its state collapses to one of the basis states $\ket{0}$ or $\ket{1}$. Hence, it is impossible to continue the algorithm after reading out. Additionally, measuring a qubit does not immediately yield the values of $\alpha$ and $\beta$ in Equation~\eqref{geng_alexander:equation_qubit}. Rather, the probability of collapsing to $\ket{0}$ is given by $|\alpha|^2$ while the state $\ket{1}$ is obtained with probability $|\beta|^2$. Repeated measurements (shots) of the same state allow for an estimation of these probabilities, and thus of the values $\alpha$ and $\beta$. For further reading on quantum computing basics, we recommend the book of Nielsen and Chuang \cite{nielsen_quantum_2010}.

For implementing our methods, we use the open-source software framework PennyLane \cite{bergholm_pennylane_2022}, which is a Python 3 \cite{van_rossum_python_2009} software framework for differentiable programming of quantum computers that enables seamless integration with machine learning tools.

\section{\label{sec:Differentiation methods}Differentiation methods}
Training a machine learning approach basically consists in solving an optimization problem. This is typically done by gradient descent methods that require the computation of derivatives. PennyLane offers in total six differentiation methods \cite{bergholm_pennylane_2022}, where we focus on numerical finite-differences (\lq finite-diff\rq), the parameter-shift rule (\lq parameter-shift\rq), and backpropagation (\lq backprop\rq). On PennyLane’s default simulator, all options are available, whereas  backpropagation cannot be used on IBM’s \qasm\, or real backends. The reason behind this is that on a real quantum computer, we do not have access to the internal computations without measuring them. Hence, intermediate results required in backpropagation are not accessible. In contrast, the parameter-shift rule and finite-differences only require evaluations of the cost function which can also be obtained on quantum computers.
More information can be found in the papers by Bergholm \textit{et al.} \cite{bergholm_pennylane_2022}, or Izaac \cite{izaac_quantum_2020}. 

Let us assume that we have some input $x\in \mathbb{R}^n$ and some parameters $\theta\in \mathbb{R}^m$, where $n,m\in \mathbb{N}$. Then, a quantum circuit can be expressed by a function
\begin{equation}\label{equ:quantum_circuit}
    f(x,\theta)\coloneqq \expect{\hat{B}}=\bra{0}U^\dagger(x,\theta)\hat{B}U(x,\theta)\ket{0},
\end{equation}
where $\hat{B}$ is an observable and $U(x,\theta)$ is the gate sequence of the quantum circuit. In the scheme of machine learning, we optimize the parameters in the training process. For that, we have to calculate the gradient of Equation~\eqref{equ:quantum_circuit}. The finite-difference method \cite{leveque_finite_2007} is based on approximating the gradient by
\begin{equation}\label{equ:finite_diff}
    \partial_\theta f(x,\theta)\approx\frac{f(x,\theta+\Delta\theta)-f(x,\theta-\Delta  \theta)}{2\Delta \theta},
\end{equation}
where $\Delta \theta$ is a small shift in the parameters. The true gradient is obtained in the limit $\Delta\theta\to 0$. 

In order to avoid this approximation, the parameter-shift rule is usually used on quantum hardware
\begin{equation}\label{equ:parameter_shift}
    \partial_\theta f(x,\theta)=c\left[f(x,\theta+s)-f(x,\theta-s)\right],
\end{equation}
where the parameters $c$ and $s$ depend on the specific function $f$ (note that only a very restricted set of functions is considered in the context of quantum computing).
It was first introduced to quantum machine learning in Mitarai et al. \cite{mitarai_quantum_2018}, and extended in Schuld et al. \cite{schuld_evaluating_2019}. Note that the shift values $s$ depend on the properties of the function  and are usually much larger than the shift values in the finite-difference method \cite{bergholm_pennylane_2022, schuld_evaluating_2019}.

The third and commonly used method for classical machine learning is backpropagation. We just need a single forward pass at the expense of increased memory usage and no further calculations. During the forward pass all intermediate steps are stored and traversed in reverse using the chain rule for adapting the parameters \cite{izaac_quantum_2020}.

The number of calls, so the number of times we have to execute the quantum circuit to obtain the results, varies significantly between the three differentiation methods.
Generally, we have 
\begin{equation}\label{equ:number_calls}
    \ncalls=\nforward+\nbackward,
\end{equation}
where $\ncalls$ is the total amount of calls needed, $\nforward$ the amount for the forward pass and $\nbackward$ the amount for the backward pass in the training. Table~\ref{tab:number_calls} shows the amount of calls required by the three methods.
\begin{table}
    \caption{\label{tab:number_calls}Number of calls $\ncalls$ needed in the training process for one epoch using three numerical methods for gradient computation. $T$ is the number of training images, $V$ the number of validation images, $Q$ the number of qubits required in the circuit for representing the features, and $L$ the number of layers (encoding and variational layers).}
    \centering
    \begin{tabular}{llll}
        \hline\noalign{\smallskip}
        Method & Forward pass & Backward pass & Total\\
        \hline
        Backpropagation     & $T+V$ & $-$                           & $T+V$\\
        Finite-difference   & $T+V$ & $T\cdot L\cdot Q$             & $T+V+T\cdot L\cdot Q$\\
        Parameter-shift     & $T+V$ & $2\cdot T\cdot L\cdot Q$      & $T+V+2\cdot T\cdot L\cdot Q$\\
        \noalign{\smallskip}\hline
    \end{tabular}
\end{table}

Consequently, every training image yields $L\cdot Q$ or $2\cdot L\cdot Q$ additional evaluations on the quantum computer for the finite-difference method or the parameter-shift rule, respectively. Depending on the number of layers $L$ and the number of qubits $Q$, this can make a huge difference, especially in the current NISQ era. 

\section{Method}\label{sec:method}
\begin{figure}[tb]
    \centering
    \includegraphics[width=\textwidth]{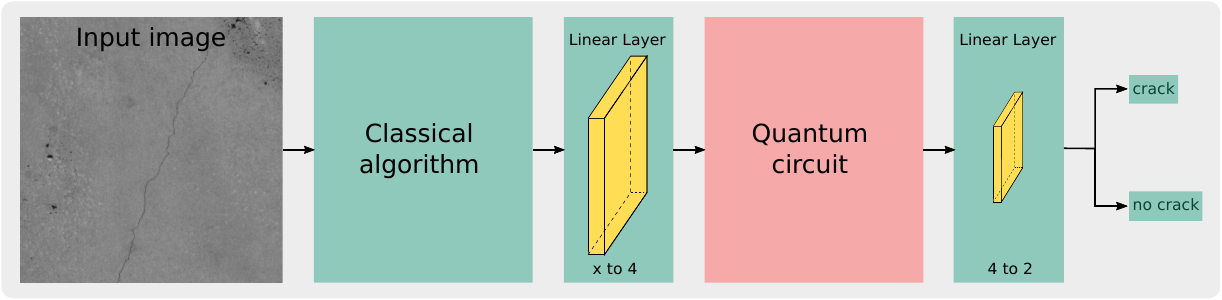}
    \caption{Schematic overview of the algorithm. The green part highlights the part for the classical computer, while the red area represents the quantum part. The variable $x$ in the first linear layer represents the number of input features. It is adaptable depending on the number of features coming from the classical algorithm. The second value is for the number of output features.}
    \label{fig:geng_alexander:schematic_overview}
\end{figure}

For solving the image classification task, we use a quantum transfer learning algorithm \cite{bergholm_pennylane_2022, mari_transfer_2020}. That means that we combine a classical neural network with some quantum part. Figure~\ref{fig:geng_alexander:schematic_overview} shows the schematic view. We insert the $224\times 224$ input images into a classical algorithm and combine the resulting features into 4 features by application of a linear layer. These four features form the input to a quantum circuit whose output is reduced from four features to two by another linear layer. Based on these two features, we can decide whether the image contains a crack or not.

\begin{figure}[tb]
    \centering
    \includegraphics[width=\textwidth]{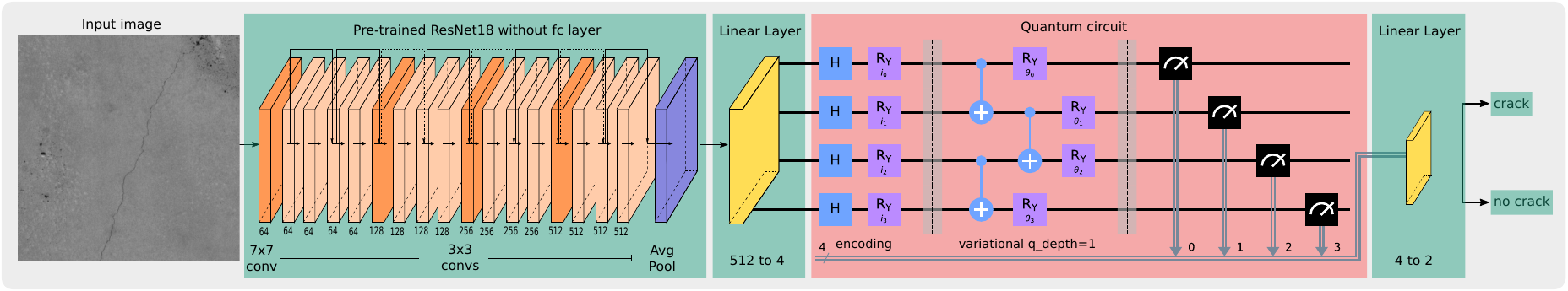}
    \caption{Proposed method for crack classification. The green part highlights the part for the classical computer, while the red area represents the quantum part. We chose ResNet18 and a quantum circuit with one variational layer.}
    \label{fig:geng_alexander:method_resnet18}
\end{figure}

In the following, we will describe the classical algorithm and the quantum circuit part in more detail. The network is visualized in Figure~\ref{fig:geng_alexander:method_resnet18}.

As classical algorithm, we use a ResNet18 network \cite{paszke_pytorch_2019} pre-trained on the ImageNet dataset \cite{russakovsky_imagenet_2015}. ResNet18 itself contains blocks of convolution layers together with batch normalization and ReLU activation layers (see the first green block in Figure~\ref{fig:geng_alexander:method_resnet18}). ResNet18 then usually applies a fully-connected layer which we replace by a hybrid method (green and red blocks after the first block). 

With the first linear layer, we reduce the number of features that we have to load into the quantum circuit. To keep the error coming from the hardware and the number of calls $\ncalls$ in Equation~\eqref{equ:number_calls} small, we decided to reduce from 512 features to 4 features in the first linear layer.
The structure of the quantum circuit follows a typical variational quantum circuit with encoding, variational, and measurement part. These parts are visualized by two barriers in the red block in Figure~\ref{fig:geng_alexander:method_resnet18}. First, we have the encoding part, also called embedding part. We start with all qubits in state $\ket{0}$. Consequently, our initial state is $\ket{0}^{\otimes Q}$, where $Q$ is the number of qubits. Afterwards, we use Hadamard gates (see Equation~\eqref{equ:gates}) to put all qubits into superposition. Then, we use \Ry-gates to encode the feature information into some rotations around the y-axis in the Bloch sphere. Hence, the number of Hadamard and \Ry-gates equals the number of qubits used for the encoding part. Here, we use a simple approach and represent each feature by one qubit. 
Theoretically, alternative encoding methods for representing the features in a quantum circuit could be used \cite{bergholm_pennylane_2022,yan_survey_2016,geng_improved_2023}.

For the variational part, there are basically two types of gates. One of them is entangling gates, usually CX-gates, the other is further rotations with trainable parameters. Here, we use a parallelized entangling scheme and \Ry-gates for the parameters. These two elements of the variational part can be repeated $q_{depth}$ times in the circuit. By that we generate more trainable parameters and also further entanglement but also more noise and more evaluations, since backpropagation is not available for adapting the parameters (see Section~\ref{sec:Differentiation methods}). If not stated otherwise, we take $q_{depth}=1$ and use  $Q=4$ qubits as visualized in Figure~\ref{fig:geng_alexander:method_resnet18} to keep the number of evaluations small. Finally, we measure all qubits in the last part of the quantum circuit.

\section{\label{sec:Application and results}Application and results}
\subsection{\label{sec:dataset}Dataset}
\begin{figure}[tb]
    \centering
    \includegraphics[width=0.6\textwidth]{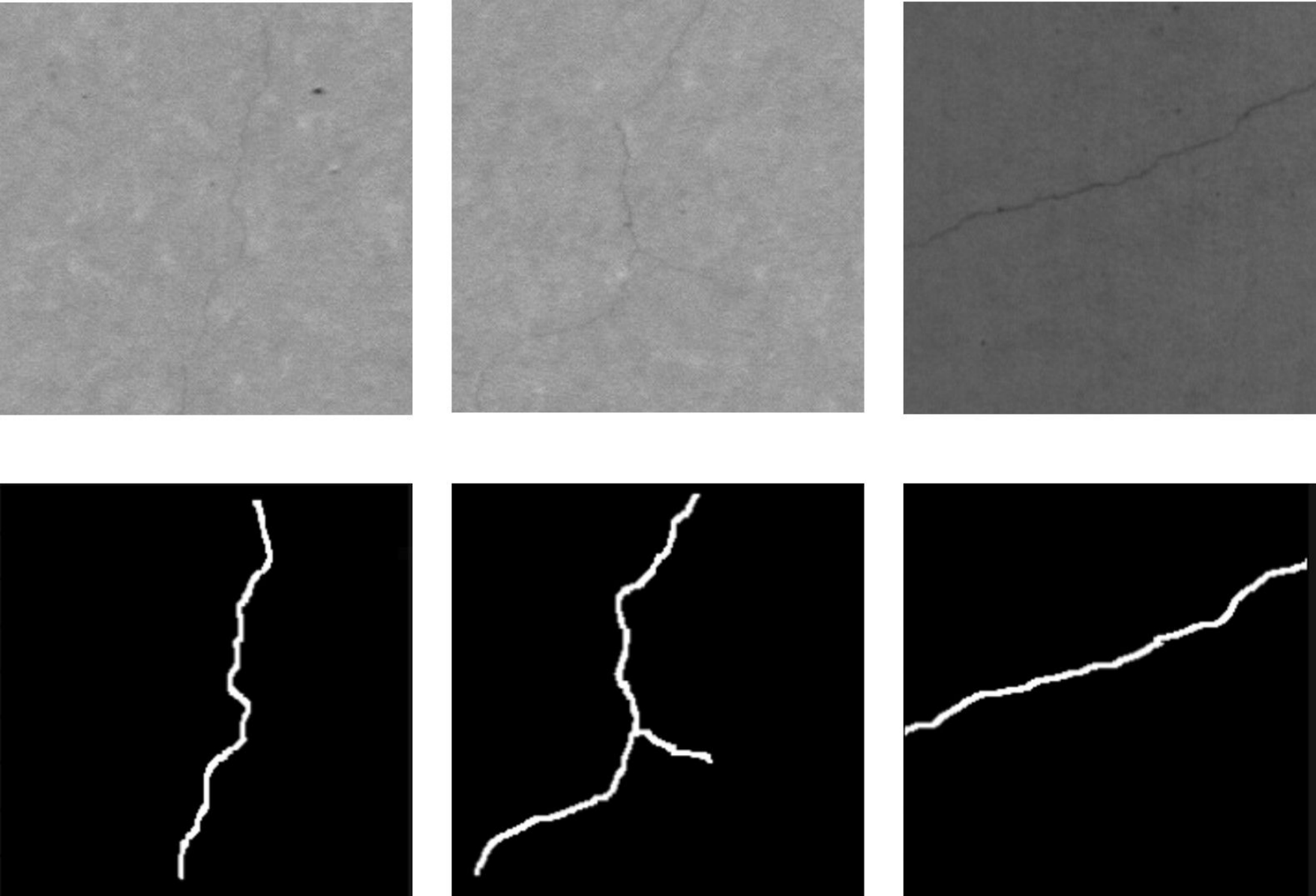}
    \caption{Sample images ($224\times 224$) of cracks with corresponding masks. The masks are purely for visualisation, they are not used for the training.}
    \label{fig:geng_alexander:sample_cracks}
\end{figure}

Our goal is to find cracks in gray value images of concrete, which have a size of around $16,000\times 32,000$ pixels. In these types of images, the cracks have a width of about 1-2 pixels. It is a challenging task to see such cracks in the images with the human eye and also some machine learning methods are not able to track them directly \cite{barisin_methods_2022}. For training, we split the images into patches of $224\times 224$ pixels and classify each patch based on whether it contains a crack or not. We used 1,223 patches as our complete dataset, 723 patches with cracks and 500 without cracks. Figure~\ref{fig:geng_alexander:sample_cracks} shows exemplary patches that contain cracks. Cracks are also visualized by masks that were obtained by manual annotation. These masks are not used for the training.

We use splits of the dataset according to the ratios $70\%/15\%/15\%$ or $4\%/4\%/92\%$ for training, validation, and testing. Table~\ref{tab:dataset_split} shows the total amount of images in the three subsets.

\begin{table}
    \caption{\label{tab:dataset_split}Splitting of the complete dataset into training, validation, and testing dataset with two different ratios.}
    \centering
    \begin{tabular}{ccccccc}
    \hline\noalign{\smallskip}
    Ratio &\multicolumn{2}{c}{Training}&\multicolumn{2}{c}{Validation} & \multicolumn{2}{c}{Testing}\\
    Train/Val/Test &cracks & no cracks&cracks & no cracks & cracks & no cracks\\ \hline
    $70\%/15\%/15\%$ & 506& 350 & 109 & 75 & 108 & 75\\
    $\phantom{1}4\%/\phantom{1}4\%/92\%$ & 29& 20 & 29 & 20 & 665 & 460\\
    \noalign{\smallskip}\hline
    \end{tabular}
\end{table}

As loss function we use cross entropy and optimize by using Adam \cite{kingma_adam_2017} with the default parameter setting. An optimization of this setting as in Geng et al.\cite{geng_comparing_2021} is beyond the scope of this paper. 

The experiments in Sections~\ref{sec:comparison} and ~\ref{sec:flexibility} are conducted on PennyLane's default simulator. In Section~\ref{sec:application_real_backend}, we use IBM's \qasm\,  and the real backends \kolkata\, and \ehningen\, and in Section~\ref{sec:application_real_backend_full_data} IBM's real backend \lima.

\subsection{\label{sec:comparison}Comparison of differentiation methods}
\begin{figure}[tb]
    \begin{subfigure}[tb]{.49\linewidth}
        \includegraphics[width=\textwidth]{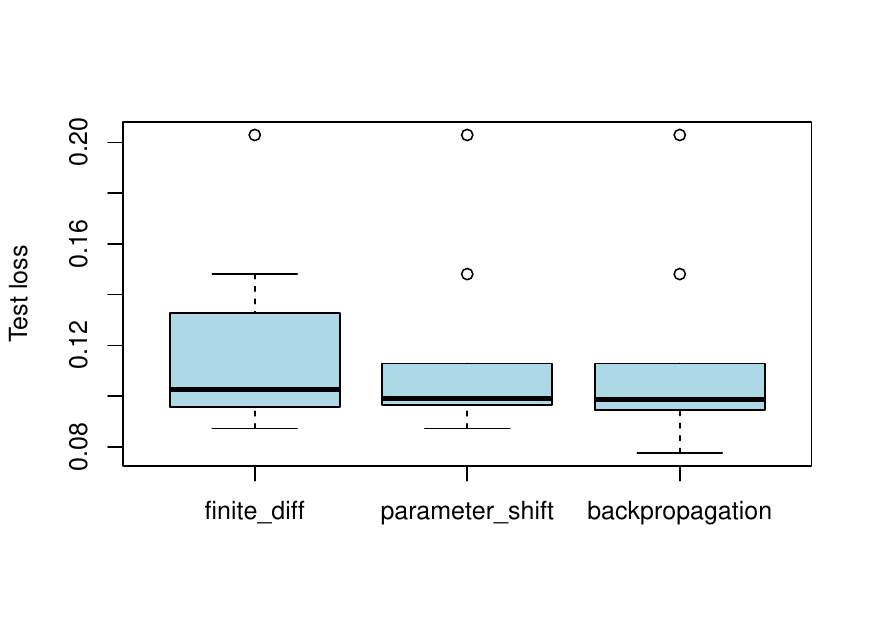}
        \caption{Loss for the three differentiation methods.}
        \label{fig:geng_alexander:loss}
    \end{subfigure}
    \begin{subfigure}[tb]{.49\linewidth}
        \includegraphics[width=\textwidth]{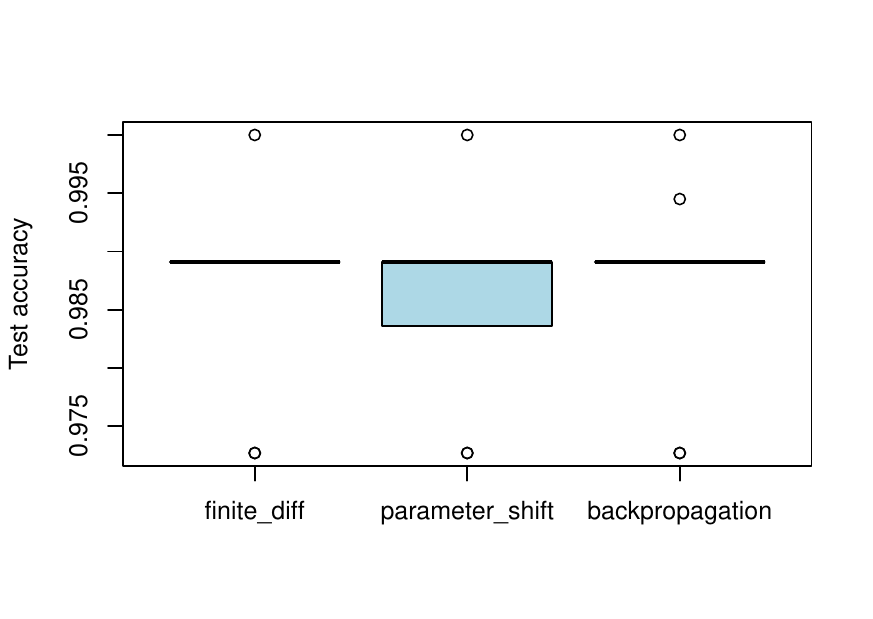}
        \caption{Accuracy for the three differentiation methods.}
        \label{fig:geng_alexander:accuracy}
    \end{subfigure}
	\caption{Loss and accuracy on the test dataset for ten runs of training with random data shuffling and 100 epochs. Performed on PennyLane's default simulator using $70\%/15\%/15\%$ splitting of the dataset for training, validation, and testing.}\label{fig:geng_alexander:loss_accuracy}
\end{figure}

Figure~\ref{fig:geng_alexander:loss_accuracy} shows the loss and the accuracy boxplots for ten runs of each of the three differentiation methods using the $70\%/15\%/15\%$ splitting of the dataset. The ten seed values are chosen randomly in the training step. Up to slight differences in the variance, the three differentiation methods yield similar results.

However, there are significant differences in terms of training time, i.e. the time needed for the $100$ epochs of training. For the calculation, we used a computer with an Intel Xeon E5-2670 processor running at $2.60$ GHz, a total RAM of $64$ GB, and Red Hat Enterprise Linux 7.9. While training with backpropagation requires in total approximately $36$ minutes, training takes much longer when using finite-differences ($\sim 45$ minutes) and the parameter-shift rule ($\sim 70$ minutes). Training times are related to the number of calls (see Table~\ref{tab:number_calls}). With $T=856$, $V=184$, $L=2$, and $Q=4$, the total number of calls per epoch is
\begin{equation}\label{equ:number_calls_1epoch}
    \ncalls=
    \begin{cases}
    \phantom{1}1,040, \text{\,backpropagation}\\
    \phantom{1}7,888, \text{\,finite-differences}\\
    14,736, \text{\,parameter-shift rule}.
    \end{cases}
\end{equation}


\subsection{\label{sec:application_real_backend}Application on two simulators and IBM's real backend}

Now we would like to compare PennyLane's default simulator with IBM's \qasm\, and IBM's real backends. To do so, we adapt the splitting of the dataset. The number of calculations required when using the splitting from the previous section presents a challenge when working on IBM's real backends. For those, fair-share queuing applies \cite{ibm}, i.e., the jobs for the different training images have to wait in the queue for being processed. In total, the waiting time exceeds the execution time. 

As backpropagation is not applicable on the real backends, we use the parameter-shift rule for all devices. To limit the required training time, we limit the number of shots to $1,000$ and the number of epochs to $10$. Thus, we have $\ncalls=8,820$. Table~\ref{tab:real_backend} shows the test losses, test accuracies, and training times for the four devices.
\begin{table}
    \caption{\label{tab:real_backend}Loss and accuracy obtained on PennyLane's default simulator, IBM's \qasm, and IBM's real backends \kolkata\, and \ehningen. Experiments on the real backend were executed on June 24 and 25, 2022. }
    \centering
    \begin{tabular}{lllr}
        \hline\noalign{\smallskip}
        Device & Test loss & Test accuracy & Training time\\
        \hline
        PennyLane simulator   & $0.5466$ & $0.7707$ & 2 min 26s\\
        \qasm               & $0.5520$ & $0.7280$ & 12 min 05s\\
        \kolkata            & $0.5245$ & $0.7102$ & 17 h 01 min 30s\\
        \ehningen           & $0.5239$ & $0.7186$ & 16 h 45 min 28s\\
        \noalign{\smallskip}\hline
    \end{tabular}
\end{table}

The higher loss and lower accuracy values compared to Figure~\ref{fig:geng_alexander:accuracy} can be explained by the lower number of training and validation images. Table~\ref{tab:real_backend} shows a higher test accuracy for PennyLane's default simulator compared to the other three methods. This is due to the absence of sampling noise coming from multiple executions and the noise introduced by the real backends. The latter ones yield quite similar results to the simulation on IBM's \qasm, which shows that the current NISQ devices are usable for this application.

Again, training time is the main issue. While the PennyLane's default simulator is optimized for quantum machine learning tasks, IBM's \qasm\, needs around six times more time to finish the calculations. The difference is mainly caused by the transition from code written in PennyLane and executed on IBM's \qasm. An even larger gap is observed between \qasm\, and IBM's backends. The differences are caused by the transition to the real device (initialization, transpilation, and validation of the circuit), but mainly by the waiting time in the queue. Execution time on the real backends is only about $2.5$ hours. Nevertheless, our experiments show that the classification of crack images is indeed achievable even on the current NISQ hardware.

\subsection{\label{sec:application_real_backend_full_data} $70\%/15\%/15\%$ dataset splitting on real backend}
\begin{figure}[tb]
    \begin{subfigure}[tb]{.51\linewidth}
        \includegraphics[width=\textwidth]{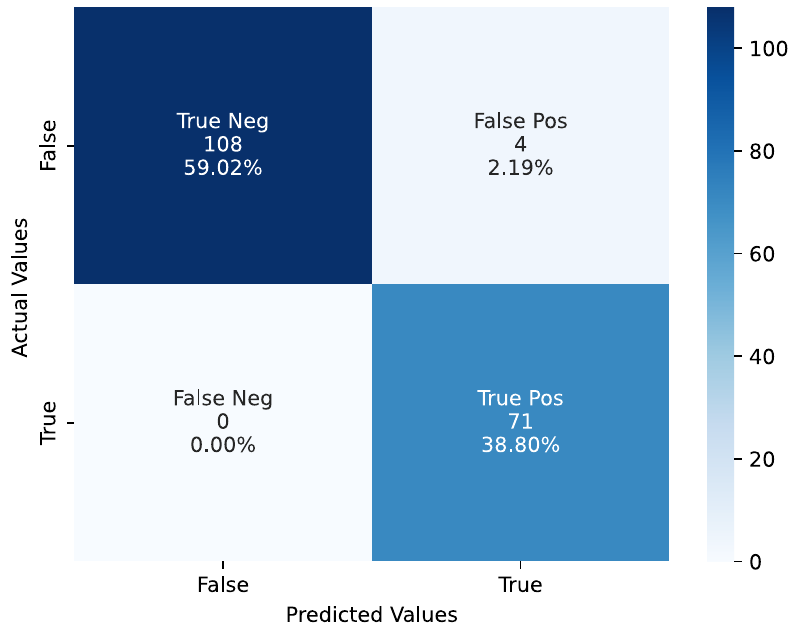}
        \caption{Confusion matrix.}
        \label{fig:geng_alexander:confusion_matrix}
    \end{subfigure}
    \begin{subfigure}[tb]{.48\linewidth}
        \begin{subfigure}{\textwidth}
            \includegraphics[width=.4\textwidth]{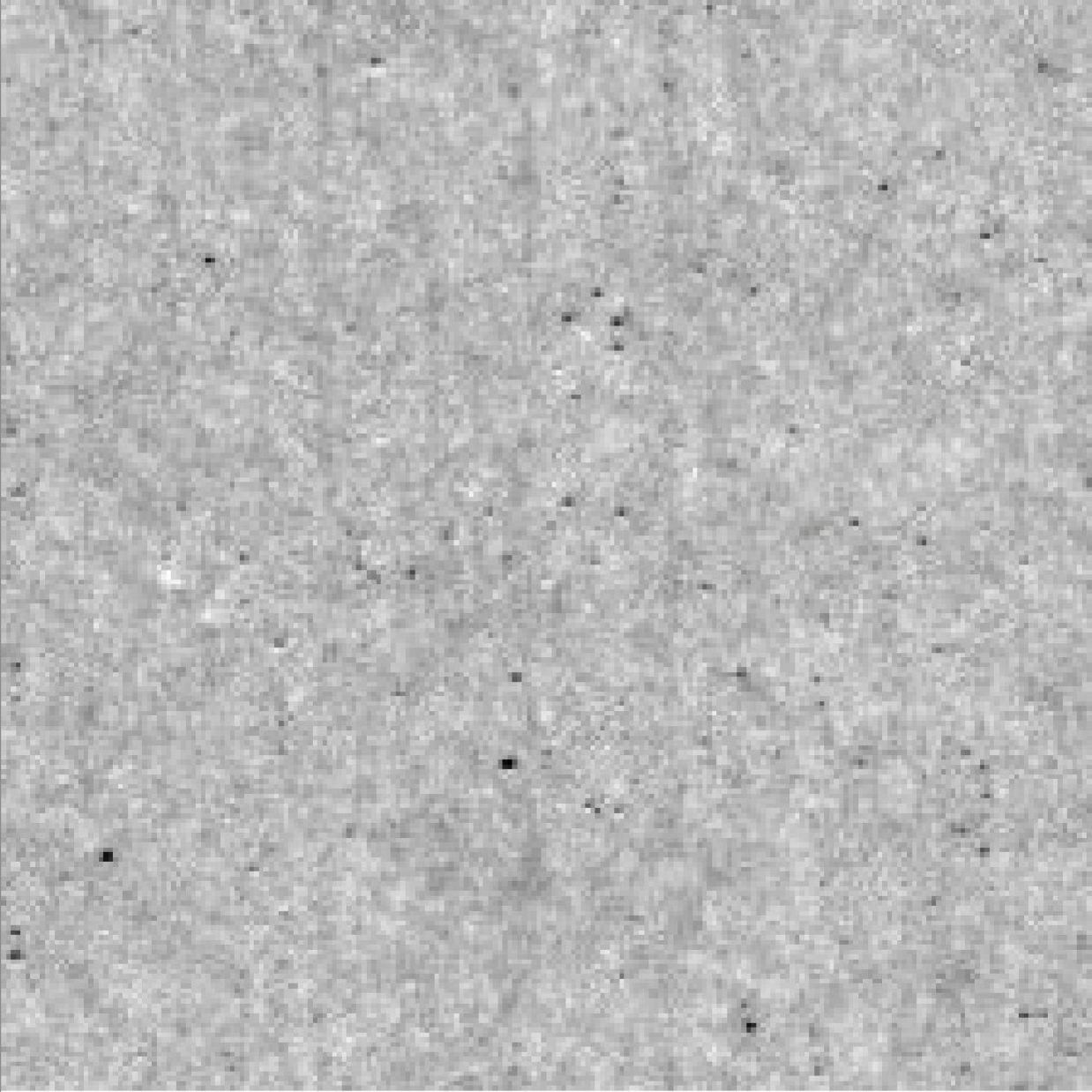}
            \includegraphics[width=.4\textwidth]{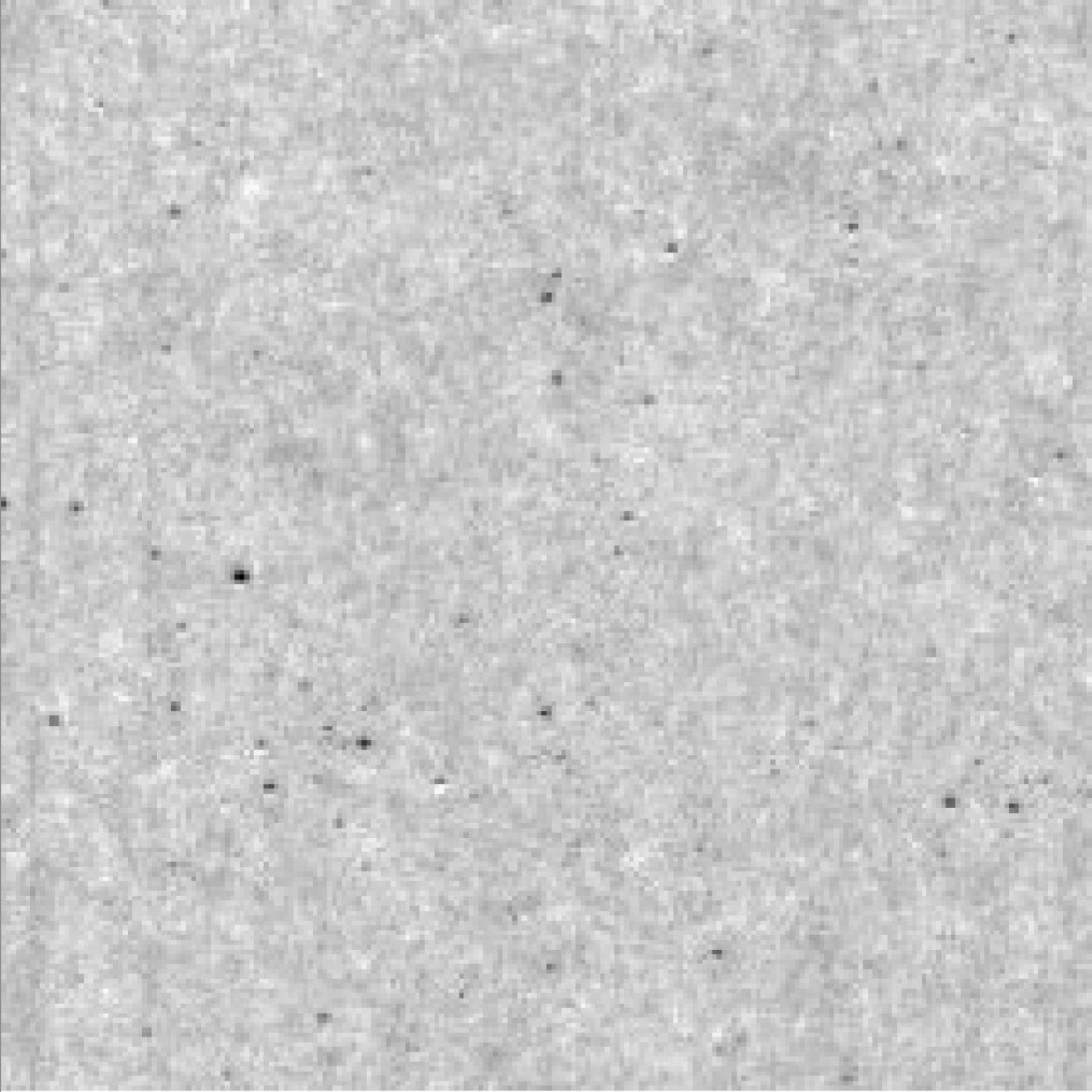}
        \end{subfigure}
        \begin{subfigure}{\textwidth}
            \includegraphics[width=.4\textwidth]{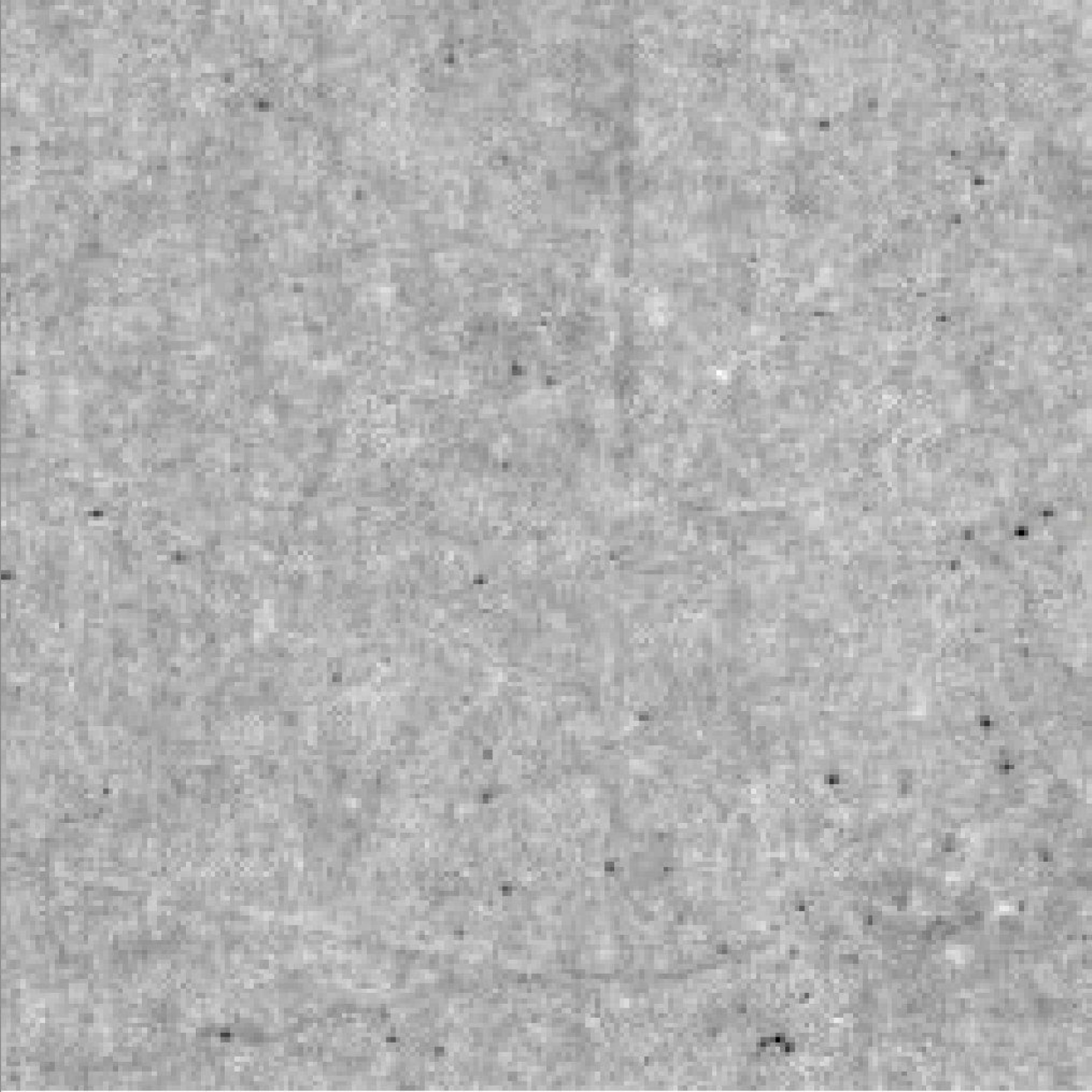}
            \includegraphics[width=.4\textwidth]{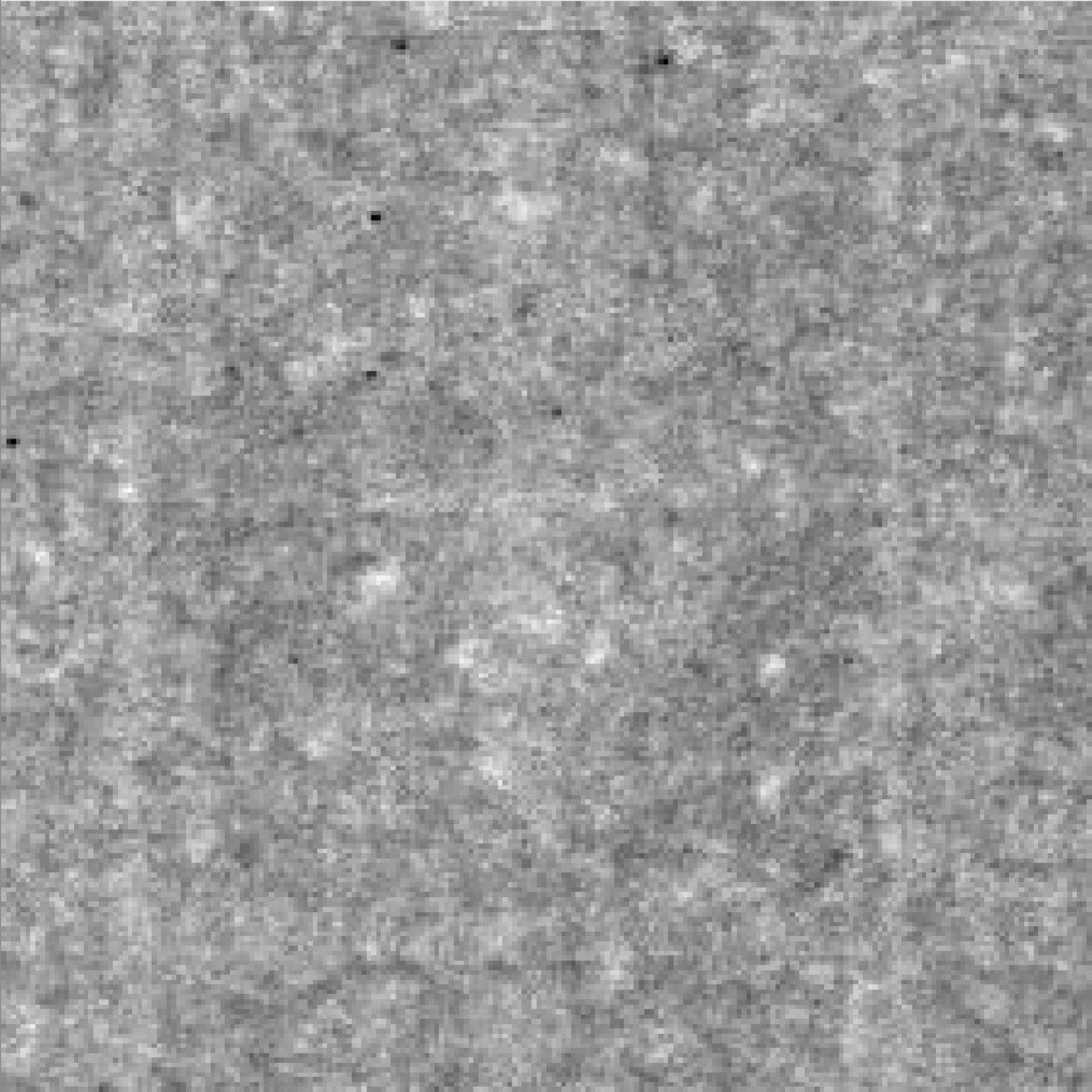}
        \end{subfigure}
        \caption{False positives.}
        \label{fig:geng_alexander:fp}
    \end{subfigure}
    \caption{Confusion matrix and false positives obtained on IBM's backend \lima. The code was executed between February 18 and 23, 2023. We have a test accuracy of $97.81\%$, which is similar to the simulator results in Figure~\ref{fig:geng_alexander:accuracy}.}
    \label{fig:geng_alexander:confusion_matrix_fp}
\end{figure}

To validate the claim that the differences in accuracies in the previous experiments are due to the different numbers of training images used, we also trained with the $70\%/15\%/15\%$ splitting on a real backend. The experiment was run on the backend \lima\, which is less frequently used than \kolkata\, and \ehningen\, such that waiting times are smaller.
Figure~\ref{fig:geng_alexander:confusion_matrix} shows the confusion matrix after one epoch training. We see that the test performance with about $97.81\%$ is comparable to the simulated results after 100 epochs (see Figure~\ref{fig:geng_alexander:accuracy}). The four false positives are visualized in Figure~\ref{fig:geng_alexander:fp}. In the images we see some pores and also some artifacts (for example, bottom left image in Figure~\ref{fig:geng_alexander:fp}) which leads to false classification. 

For the training, we needed multiple executions on the real backend (in total $14,736$, see Equation~\eqref{equ:number_calls_1epoch}), which yields to a total training time of about 4 days 6 hours and 30 minutes for one epoch. As already stated in the previous section, the main amount of time we wait in the queue. The actual execution time on \lima\, was 4 hours, so only about $4\%$ of the total training time. In comparison, a classical deep learning approach would only need 3 hours for 100 epochs of training.

\section{\label{sec:flexibility}Modifications of the algorithm}
\begin{figure}[tb]
    \centering
    \includegraphics[width=\textwidth]{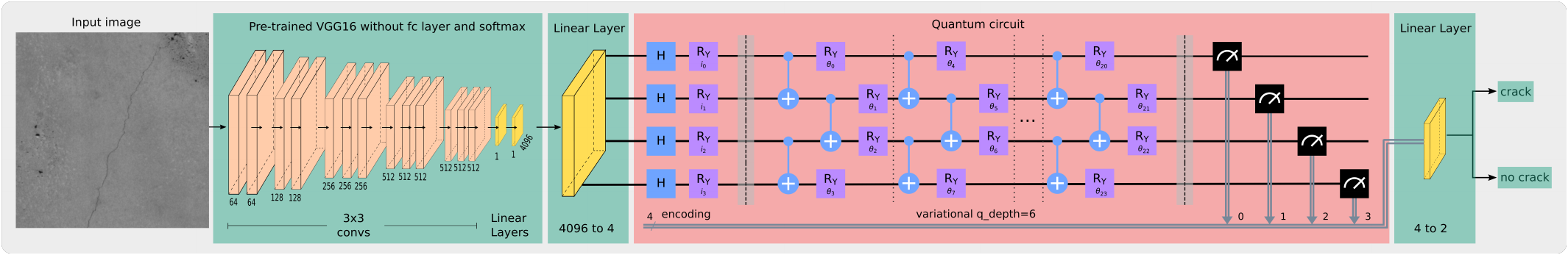}
    \caption{Alternative method for crack classification. The green part highlights the part for the classical computer, while the red area represents the quantum part. We chose the VGG16 network and a quantum circuit with six variational layers (three of which are drawn here).}
    \label{fig:geng_alexander:method_vgg16}
\end{figure}

There are various options to modify our approach. One option is to choose an alternative classical network to replace the existing classical part, like, for example, VGG16 \cite{simonyan_very_2015} instead of ResNet18. The green part in Figure~\ref{fig:geng_alexander:method_vgg16} shows the scheme of the alternative method using VGG16. Again, the output of the network, in this case $4,096$ features, is reduced to 4 features. By that, we have then the same input for the quantum circuit as in the previous section. 

Another option is to change the quantum circuit by using another entanglement strategy (like all-to-all, linear, or circular entanglement), another encoding method (like amplitude, or angle encoding \cite{geng_improved_2023}), or a higher $q_{depth}$ value. We cover the last option in this paper by varying it between 1 and 6. The red framed part in Figure~\ref{fig:geng_alexander:method_vgg16} shows the scheme of a possible quantum circuit with $q_{depth}=6$ and 24 trainable parameters. For the options with changing entanglement or encoding, we refer to Aleksandrowicz \textit{et al.}\cite{aleksandrowicz_qiskit_2019} and Bergholm \textit{et al.} \cite{bergholm_pennylane_2022}.

\begin{figure}[tb]
    \begin{subfigure}[tb]{.49\linewidth}
        \includegraphics[width=\textwidth]{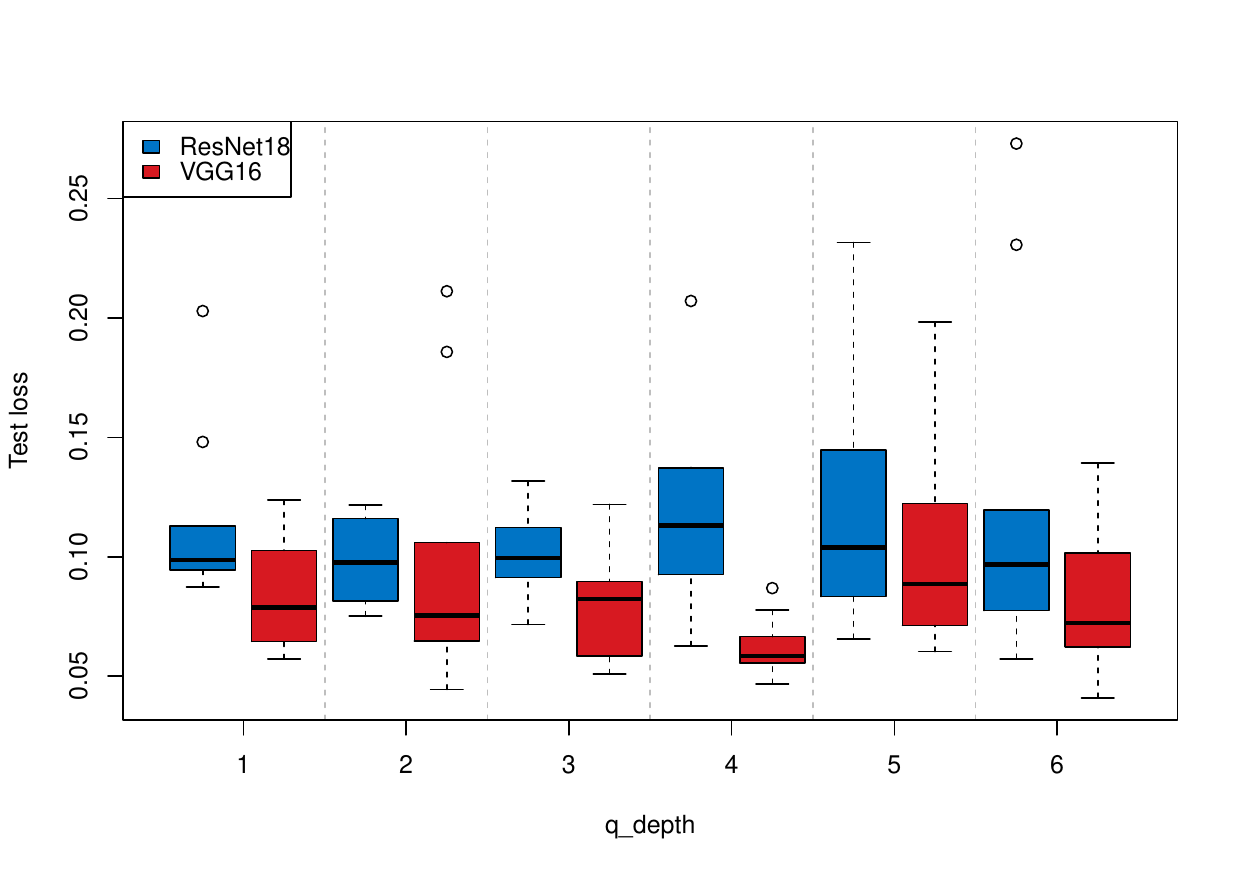}
        \caption{Loss values.}
        \label{fig:geng_alexander:loss_vgg16}
    \end{subfigure}
    \begin{subfigure}[tb]{.49\linewidth}
        \includegraphics[width=\textwidth]{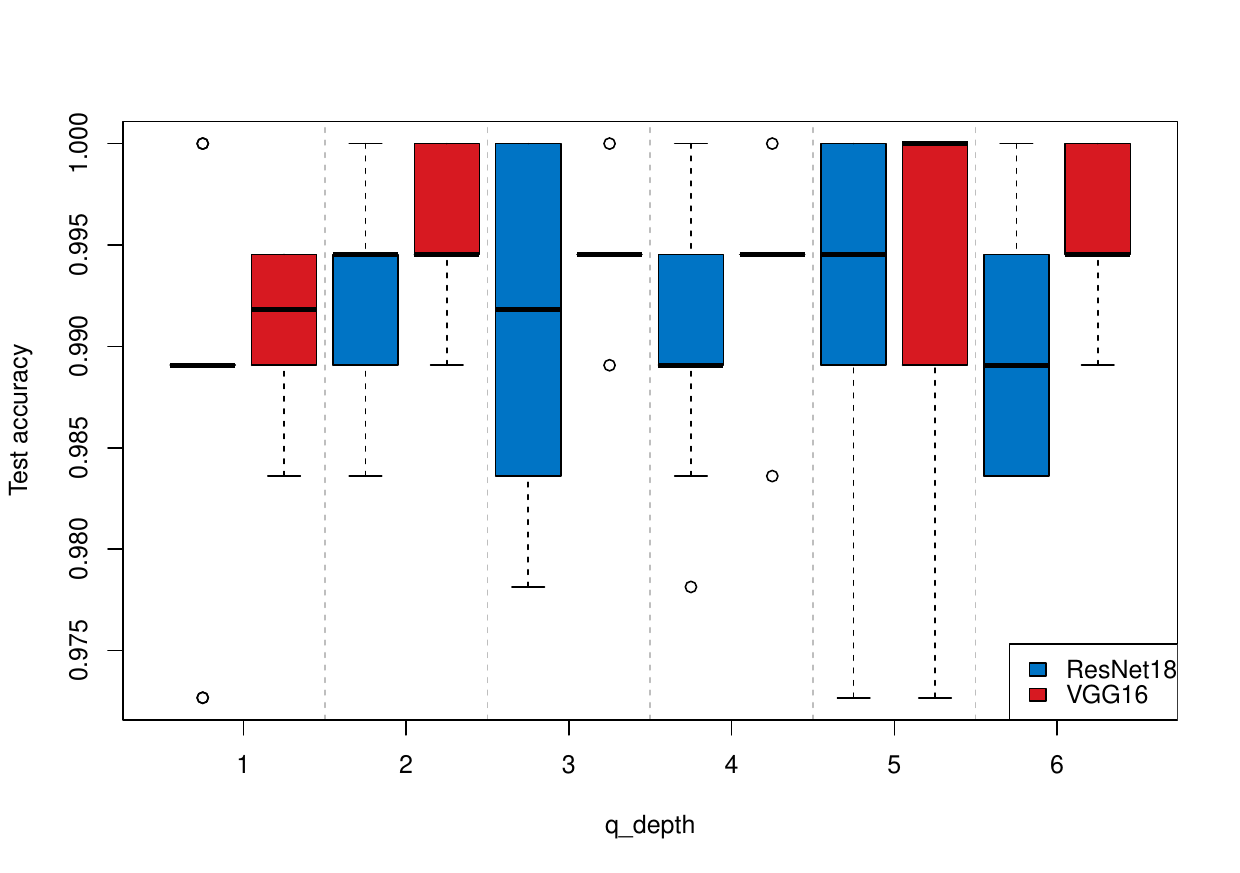}
        \caption{Accuracy value.}
        \label{fig:geng_alexander:accuracy_vgg16}
    \end{subfigure}
	\caption{Loss and accuracy obtained for training on PennyLane's default simulator by using backpropagation. Results of ten runs with random data shuffling, ResNet18 and VGG16 as pre-trained classical networks, and varying number of variational layers $q_{depth}\in\{1,2,\dots,6\}$.}
	\label{fig:geng_alexander:loss_accuracy_vgg16}
\end{figure}

In all cases, we use the $70\%/15\%/15\%$ splitting of the dataset. Figure~\ref{fig:geng_alexander:loss_accuracy_vgg16} shows that changing the classical network affects the results slightly. 
In terms of variational layers, we see a minimal improvement when using more layers. However, this comes at the price of a larger number of calls. 
Especially, for finite-difference and parameter-shift rule, this makes a difference since the number of variational layers $L=q_{depth}+1$ is one factor for the backward pass (see Table~\ref{tab:number_calls}).

Table~\ref{tab:execution_times_vgg16_resnet18} compares the training times using ResNet18 and VGG16 with $q_{depth}\in\{1,2,\dots,6\}$. The training time only differs slightly over the ten runs, which is why we just show the mean value here. Since VGG16 is bigger than ResNet18 in terms of parameters and in our case the first linear layer decreases the number of features from $4,096$ to $4$ instead of $512$ to $4$, we see a higher training time for VGG16. By increasing the number of variational layers, we have a linear increase in the training times.
\begin{table}
\caption{\label{tab:execution_times_vgg16_resnet18}Mean training times for the results shown in Figure~\ref{fig:geng_alexander:loss_accuracy_vgg16}.}
\centering
\begin{tabular}{lcccccc}
\hline\noalign{\smallskip}
Pre-trained &\multicolumn{6}{c}{$q_{depth}$}\\
network     & 1 &   2   &   3   &   4   &   5   &   6\\
\hline
ResNet18     & 36 min 20s & 44 min 28s & 55 min 22s &65 min 55s & 77 min 28s& 83 min 08s\\
VGG16       & 37 min 53s & 49 min 39s & 60 min 10s &68 min 34s & 82 min 16s& 90 min 22s\\
\noalign{\smallskip}\hline
\end{tabular}
\end{table}

\section{\label{sec:Conclusion and Outlook}Conclusion and Outlook}
In this paper, we demonstrate that the task of crack detection in gray value images can be solved by using current IBM's superconducting quantum computers. At the moment, backpropagation is not available on real devices. So, using finite-difference or parameter-shift rule are the options currently used to calculate the gradients. A larger number of evaluations, longer training times, and more jobs to submit are the consequences. In general, we achieved similar results on PennyLane's and IBM's simulators and IBM's superconducting quantum computers. Once trained, it is possible to use the trained models for real data from an industrial setting.

Our experiments show that the architecture of the quantum circuit is an important factor for the execution time. The more parameters we have within the variational part, the more time is needed for the backward pass in the quantum machine learning setting. However, we demonstrate, that increasing the number of parameters by using more variational layers does not automatically improve the results.

Our proposed algorithm can be adapted to other use-cases. Quantum transfer learning is not limited to 2D gray value images and we can also change it to 3D images as those processed by Barisin \textit{et al.} \cite{barisin_methods_2022} in a classical regime. Currently, it seems to be more promising to change the classical part to adapt the algorithm to 3D images, but with further improvement of the quantum hardware, a higher proportion of the algorithm could be run in the quantum part.

Since classical machine learning benefits from the fast gradient calculation using backpropagation, optimizing gradient computation is in the focus of current quantum computing research. With structured quantum circuits, the gradient estimation requires fewer circuits \cite{bowles_backpropagation_2023}. Looking more into this direction and also adapting the variational ansatz to it, could be a promising way to overcome the drawback of calculating as many circuits on the quantum hardware as done in this paper.

\newpage
\bibliographystyle{unsrt}
\bibliography{main}

\newpage
\section{Declarations}
\subsection{Funding}
This work was supported by the project AnQuC-3 of the Competence Center Quantum Computing Rhineland-Palatinate (Germany) and by the German Federal Ministry of Education and Research (BMBF) under grant 05M2020 (DAnoBi). Additionally, this work was funded by the Federal Ministry for Economic Affairs and Climate Action (German: Bundesministerium für Wirtschaft und Klimaschutz) under the project EniQmA with funding number 01MQ22007A.
\subsection{Conflicts of interest/Competing interests}
The authors declare no competing interests.
\subsection{Availability of data and material}
All the data and simulations that support the findings are available from the corresponding author on request. 
\subsection{Code availability}
The jupyter notebooks used in this study are available from the corresponding author on request.
\subsection{Ethics approval}
\subsection{Consent to participate}
\subsection{Consent for publication}

\end{document}